# Belykh attractor in Zaslavsky map and its transformation under smoothing

*S.P. Kuznetsov*

Kotelnikov's Institute of Radio-Engineering and Electronics of RAS, Saratov Branch
Zelenaya 38, Saratov, 410019, Russia

*spkuz@yandex.ru*

If we allow non-smooth or discontinuous functions in definition of an evolution operator for dynamical systems, then situations of quasi-hyperbolic chaotic dynamics often occur like, for example, on attractors in model Lozi map and in Belykh map. The present article deals with the quasihyperbolic attractor of Belykh in a map describing a rotator with dissipation driven by periodic kicks, the intensity of which depends on the instantaneous angular coordinate of the rotator as a sawtooth-like function, and also the transformation of the attractor under smoothing of that function is considered. Reduction of the equations to the standard form of the Belykh map is provided. Results of computations illustrating the dynamics of the system with continuous time on the Belykh attractor are presented. Also, results for the model with the smoothed sawtooth function are considered depending on the parameter characterizing the smoothing scale. On graphs of Lyapunov exponents versus a parameter, the smoothing of the sawtooth implies appearance of periodicity windows, which indicates violation of the quasi-hyperbolic nature of the attractor. Charts of dynamic regimes on the parameter plane of the system are also plotted, where regions of periodic motions ("Arnold's tongues") are present, which decrease in size with the decrease in the characteristic scale of the smoothing, and disappear in the limit case of the sawtooth function with a break. Since the Belykh attractor was originally introduced in the radiophysical context (phase-locked loops), the analysis undertaken here is of interest from the point of view of possible exploiting the chaotic dynamics on this attractor in electronic devices.

## Introduction

From the point of view of possible applications of chaos, it is of great importance that at a small perturbation of the system parameters the generated chaos would not degrade and be transformed into modes of regular, say, periodic dynamics [1]. If we talk about systems where the evolution operator is represented with smooth functions of state variables, then this kind of "genuine" robust chaos is realized for strange attractors of hyperbolic type (Smale-Williams attractor, Plykin attractor) possessing a mathematically grounded property of roughness or structural stability [2-4], and for pseudohyperbolic attractors (the Lorenz attractor, "wild" attractors) [5-7]. However, much more often in the smooth case one has to deal with a situation of a quasiatractor (the Hénon map, the Rössler model, many examples in electronics and other disciplines) [8-13]. In the case of quasiatractor, the dynamics observed in numerical calculations or in experiments, although it look chaotic, are characterized by the presence in the same region of the phase space of regular attracting sets ("sinks") with narrow basins of attraction, or they arise with arbitrarily small variation of the system parameters.

On the other hand, if the evolution operator of the dynamical system is constructed allowing non-smooth or discontinuous functions, robust quasi-hyperbolic chaos is often easily attainable [12-16]. Examples are the Lozi attractor in the model two-dimensional map given by a piecewise linear function [17, 18] and the Belykh attractor, which was introduced some time ago in the context of phase-locked loops [19], under the assumption of a sawtooth form of the dependence of the transfer function on the initial phase [20-22]. Such situations correspond to chaotic dynamics in many respects analogous to hyperbolic attractors (with the exception of singular trajectories visiting the points of discontinuity, which can be ignored as atypical ones).

At the same time, when the functions appearing in the definition of the evolution operator are smoothed out, the quasi-hyperbolic nature of the chaos is violated usually, and the dynamic behavior corresponding to quasiatractors, arises; in particular, "windows of regularity" appear that are distinguishable with variations of the parameters. For models in the form of one-dimensional maps and for the Lozi attractor, this point has been discussed in the literature, see, for example, [23,24].

The present paper is devoted to consideration of the Belykh attractor and to discussion of its transformation when smoothing the sawtooth function used in the model. Since the Belykh attractor originally appeared in the context of radiophysical problems, the analysis undertaken is of interest from the point of view of possible use of chaotic dynamics on this attractor in electronic devices. In Section 1 we consider a concrete example of a dissipative system where this type of dynamics takes place, namely, a rotator under the action of periodic kicks. The description of the state change on a period of kicks leads to the Zaslavsky map [25, 26], in which, when choosing the sawtooth form with a break for the function of intensity of the kicks on the rotation angle $g(\theta)$, the Belykh attractor occurs. Section 2 presents the corresponding analytic derivations reducing the problem to the standard form of the Belykh map [22]. In Section 3 we give illustrations of the dynamics of the system with continuous time on the Belykh attractor. Section 4 discusses the dynamics of a model with a smoothed $g(\theta)$ as a function of a parameter characterizing the smoothing scale. In particular, graphs of Lyapunov exponents are presented, on which, as the $g(\theta)$ function is smoothed, more and more distinct windows of periodic dynamics appear. Also charts of regimes on parameter plane are presented illustrating transformation of the picture of synchronization tongues and chaotic domains.

## 1. Zaslavsky map and kicked rotator

Zaslavsky map describes dynamics of many dissipative systems characterized by periodicity of the state space with respect to one of the variables. They include a rotator under periodic kicks [25, 26], pulsed phase phase-locked loops [19], a problem of dynamics near a limit cycle with periodic kicks [4].

Following [25, 26], we consider a rotator characterized by the angular coordinate $\theta$ and the angular velocity $\omega$ under periodic pulsed kicks, whose intensity depends on the instantaneous angular coordinate by means of a function that has a period of $2\pi$ in its argument. Assuming that a constant torque $M$ is also applied, the equations can be written in the form

$$\dot{\theta} = \omega, \quad J\dot{\omega} = -\gamma\omega + kg(\theta)\sum_{n=-\infty}^{\infty}\delta(t-nT) + M, \qquad (1)$$

where $J$ is moment of inertia, $\gamma$ is a coefficient of friction, and $k$ is a parameter characterizing the intensity of the kicks. We introduce a dimensionless time $\tau = t/T$ and a dimensionless angular relative velocity $w = (\omega - M\gamma^{-1})T$, and then we have

$$\frac{d\theta}{d\tau} = w + \Delta, \quad \frac{dw}{d\tau} = -\Gamma w + Kg(\theta)\sum_{n=-\infty}^{\infty}\delta(\tau-n), \qquad (2)$$

where $\Gamma = \gamma T J^{-1}$, $K = kTJ^{-1}$, $\Delta = MTJ^{-1}$.

Suppose that at the moment just before the $n$-th pulse the state of the rotator is given by variables $(\theta_n, w_n)$. Immediately after the kick, the values of the variables will be $(\theta_n, w_n + Kg(\theta_n))$. Solving the equations describing the motion in the interval between the kicks with these initial conditions, we have:

$$\begin{aligned}w_{n+1} &= [w_n + Kg(\theta_n)]e^{-\Gamma}, \\ \theta_{n+1} &= \theta_n + [w_n + Kg(\theta_n)]\frac{1-e^{-\Gamma}}{\Gamma} + \Delta.\end{aligned} \qquad (3)$$

Introducing the notation $x_n = \theta_n$, $y_n = \omega_n \frac{1-e^{-\Gamma}}{\Gamma}$, $a = K\frac{1-e^{-\Gamma}}{\Gamma}$, $\lambda = e^{-\Gamma}$, we arrive at the following compact form of the Zaslavsky map

$$x_{n+1} = x_n + y_n + ag(x_n) + \Delta \pmod{2\pi}, \quad y_{n+1} = \lambda[y_n + ag(x_n)]. \tag{4}$$

Calculation of the Jacobi determinant

$$\begin{vmatrix} \partial x_{n+1}/\partial x_n & \partial x_{n+1}/\partial y_n \\ \partial y_{n+1}/\partial x_n & \partial y_{n+1}/\partial y_n \end{vmatrix} = \begin{vmatrix} 1+ag'(x_n) & 1 \\ \lambda g'(x_n) & \lambda \end{vmatrix} = \lambda \tag{5}$$

shows that in the parameter region $0 < \lambda < 1$ the map is area compressing, so that attractors must be present.

## 2. Belykh attractor

If the dependence $g(x)$ in the equations (1)-(4) is defined as a discontinuous function shown in Fig. 1,

$$g(x) = \begin{cases} x/\pi + 1, & -\pi < x < 0, \\ x/\pi - 1, & 0 < x < \pi, \end{cases} \quad g(x+2\pi) = g(x), \tag{6}$$

then the map (4) has a quasi-hyperbolic attractor in a certain parameter domain. This attractor was originally introduced and studied by V.N. Belykh in the context of phase-locked loops [20-22].

Figure 2a shows attractor of the map (4) with the function (6) and with values of the parameters

$$a=1.2, \Delta=0, \lambda=0.48. \tag{7}$$

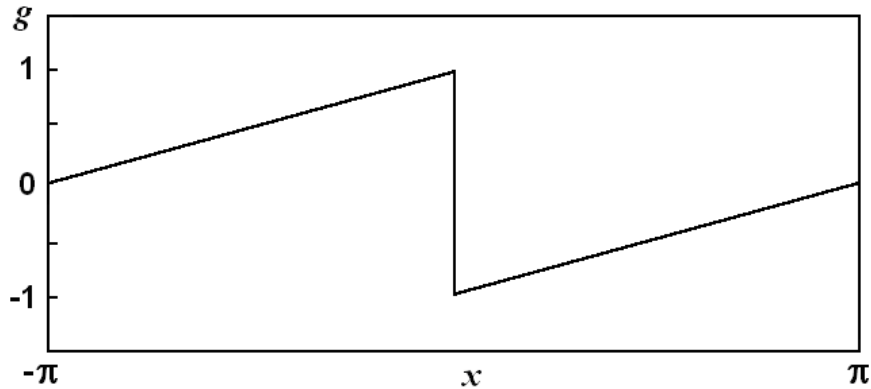

Fig.1. Plot of the discontinuous sawtooth function $g(x)$ used in the definition of the evolution operator for the system with Belykh attractor.

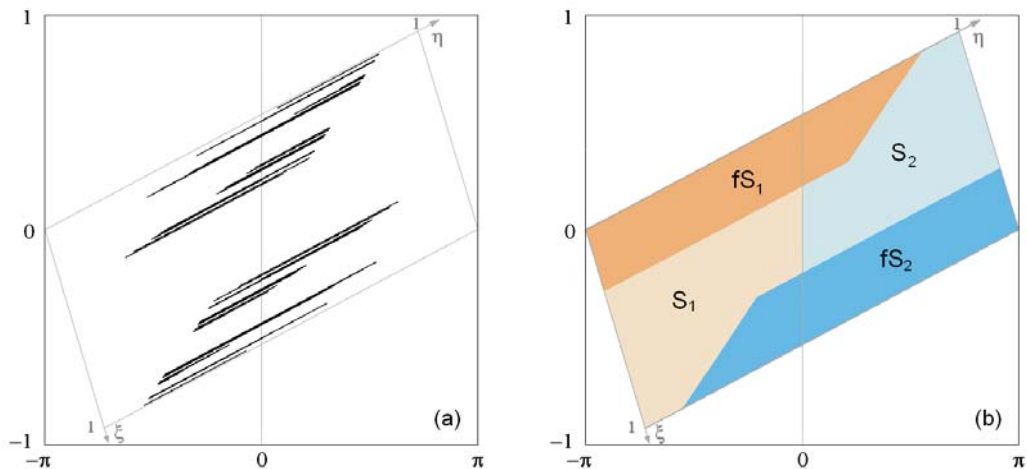

Fig.2. Attractor of the map (4) with the function (6) for $a=1.2, \Delta=0, \lambda=0.48$ (a) and configuration of the areas arising when the map is applied (b).

To reproduce the construction of Belykh, we note that the Jacobi matrix corresponding to the map (4) with the function (6) is

$$\begin{pmatrix} 1+a/\pi & 1 \\ \lambda a/\pi & \lambda \end{pmatrix}, \qquad (8)$$

and its eigenvalues and eigenvectors are expressed as follows:

$$\lambda_1 = \frac{a+\pi+\pi\lambda - \sqrt{(a+\pi+\pi\lambda)^2 - 4\pi^2\lambda}}{2\pi}, \quad \lambda_2 = \frac{a+\pi+\pi\lambda + \sqrt{(a+\pi+\pi\lambda)^2 - 4\pi^2\lambda}}{2\pi}, \qquad (9)$$

$$\begin{pmatrix} u_1 \\ v_1 \end{pmatrix} = \frac{\pi}{\sqrt{(a+\pi+\pi\lambda)^2 - 4\pi^2\lambda}} \begin{pmatrix} -a-\pi+\pi\lambda + \sqrt{(a+\pi+\pi\lambda)^2 - 4\pi^2\lambda} \\ -2\lambda a \end{pmatrix}, \qquad (10)$$

$$\begin{pmatrix} u_2 \\ v_2 \end{pmatrix} = \frac{\pi}{\sqrt{(a+\pi+\pi\lambda)^2 - 4\pi^2\lambda}} \begin{pmatrix} a+\pi-\pi\lambda + \sqrt{(a+\pi+\pi\lambda)^2 - 4\pi^2\lambda} \\ 2\lambda a \end{pmatrix}. \qquad (11)$$

The normalization of the eigenvectors is chosen here in such way that

$$u_1 + u_2 = 2\pi, \; v_1 + v_2 = 0.$$

The map (4) has a fixed point $x_0 = -\pi - \pi(1-\lambda)\Delta/a$, $y_0 = -\lambda\Delta$, which we take as an origin of the new coordinate system $(\xi, \eta)$, defining its axes as directions of the eigenvectors. In Fig. 2, the axes of the new coordinate system are indicated as gray straight lines. The connection of new variables with the initial coordinates is given by the relations

$$x = -\pi - \pi(1-\lambda)\Delta/a + \xi u_1 + \eta u_2, \; y = -\lambda\Delta + \xi v_1 + \eta v_2, \qquad (12)$$

or by the reverse variable change

$$\xi = \frac{[x+\pi+\pi(1-\lambda)\Delta/a]v_2 - (y+\lambda\Delta)u_2}{u_1 v_2 - u_2 v_1}, \; \eta = \frac{-[x+\pi+\pi(1-\lambda)\Delta/a]v_1 + (y+\lambda\Delta)u_1}{u_1 v_2 - u_2 v_1}. \qquad (13)$$

Consider a region in a form of quadrangle whose two vertices $(-\pi, 0)$ and $(\pi, 0)$ correspond to fixed points of the map, and the sides are parallel to the axes $\xi$ and $\eta$. This region is divided into two parts $S_1$ and $S_2$ in Fig.2 by the vertical line $x=0$, on which the map has a discontinuity. The subdomains $S_1$ and $S_2$ are mapped into the region in the form of two trapezes $fS_1$ and $fS_2$ stretched out along an unstable eigendirection and compressed along a stable direction (Fig. 2b).

Figure 3 shows similar images, but in the eigenvector coordinates $(\xi, \eta)$, which correspond to the pictures for Belykh map in the standard form given in [22].

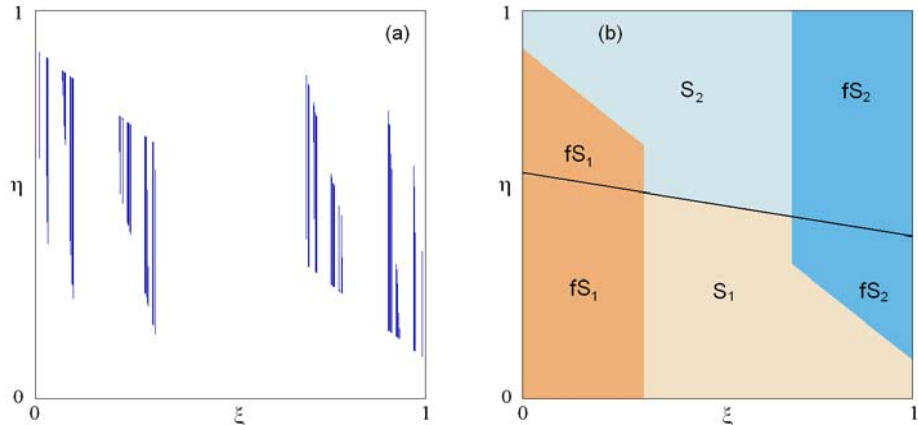

Fig.3. Attractor of the map (4) with the function (6) for $a=1.2$, $\Delta=0$, $\lambda=0.48$ (a) and configuration of the regions arising when applying the map (b) in coordinates defined through the eigenvectors. The inclined black line shows the disposition of the discontinuity and is given by the equation $(\xi - \tfrac{1}{2})u_1 + (\eta - \tfrac{1}{2})u_2 = 0$.

The Lyapunov exponents calculated for this attractor using the standard algorithm [27, 28, 14] are

$$\Lambda_1 = 0.440, \Lambda_2 = -1.174. \qquad (12)$$

Note that these quantities are exactly logarithms of the eigenvalues (9) of the matrix (8), as it is natural because of the piecewise-linear nature of the mapping. The dimension of the attractor by the Kaplan – Yorke formula [29, 4] is $D_{KI} = 1 + \Lambda_1 / |\Lambda_2| = 1.375$, i.e. it is fractional; indeed, the attractor surely is a fractal object as seen from the images in Fig. 2a and 3a.

## 3. Attractor of the system with continuous time

Let us return to equations of motion of the rotator with dissipation under the action of pulsed kicks as a system with continuous time (2). Figure 4 shows the graphs of the time dependences of the angular coordinate and the dimensionless angular velocity in the course of the dynamics on the Belykh attractor, which is realized when the system parameters are assigned in accordance to the parameters of the map (6). Figure 5 shows the attractor of the continuous time system (2) in the extended phase space with the same parameters. Due to the periodicity of the extended phase space over the time coordinate, the picture has to be imagined as repetitive with period 1 along the vertical axis.

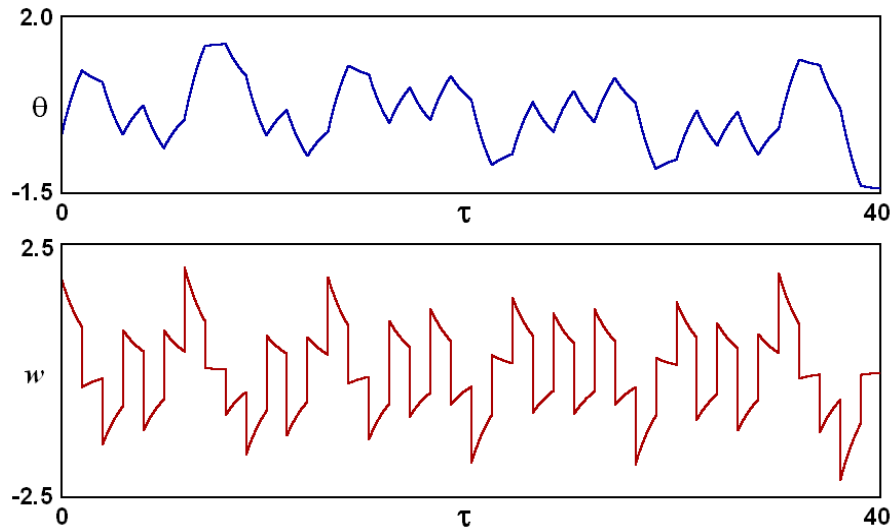

Fig.4. Time dependencies of the angular coordinate and the dimensionless angular velocity for dynamics on the Belykh attractor of the continuous time system (2); the values of parameters $\Gamma = 0.7340$, $K = 1.6938$, $\Delta = 0$ correspond to the parameters of the map (6) $a=1.2$, $\lambda=0.48$.

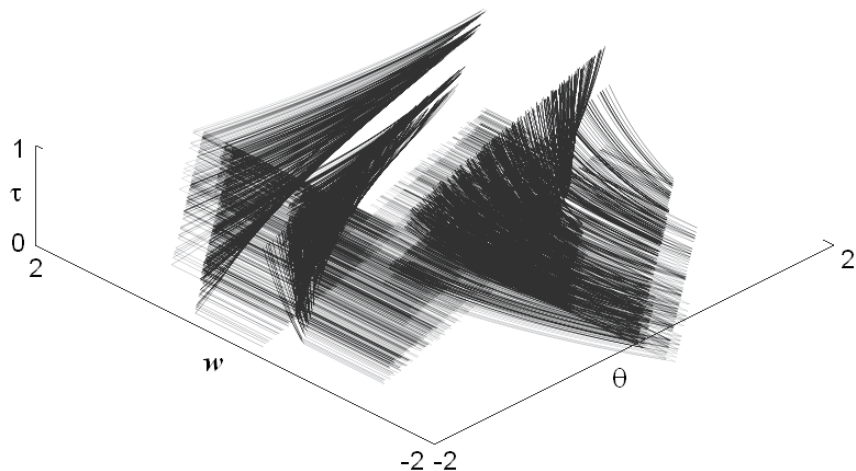

Fig.5. Belykh attractor of the system with continuous time (2) in the extended phase space; the image should be regarded as periodically repetitive with period 1 along the vertical axis. The values of the parameters $\Gamma = 0.7340$, $K = 1.6938$, $\Delta = 0$ correspond to the parameters of the map (6) $a=1.2$, $\lambda=0.48$.

## 4. The model with smoothing: Destruction of the Belykh attractor

Now let us turn to discussion of how the dynamics on the given type of attractor varies when the sawtooth function in the definition of the evolution operator undergoes smoothing. To do this, we introduce the following family of functions depending on a parameter µ (Fig. 6):

$$g_\mu(x) = \begin{cases} \frac{1}{2}\alpha^3 x^3 - \frac{3}{2}\alpha x, & |\alpha x| \leq \mu+1, \\ (1+\mu)^3 (x/\pi - \operatorname{sgn} x), & |\alpha x| > \mu+1, \end{cases} \qquad (14)$$

$$\alpha = \frac{2(1+\mu)^3}{3\pi\mu(2+\mu)}, \quad g(x+2\pi) = g(x).$$

The limit µ→ 0 corresponds to the discontinuous function shown in Fig. 1.

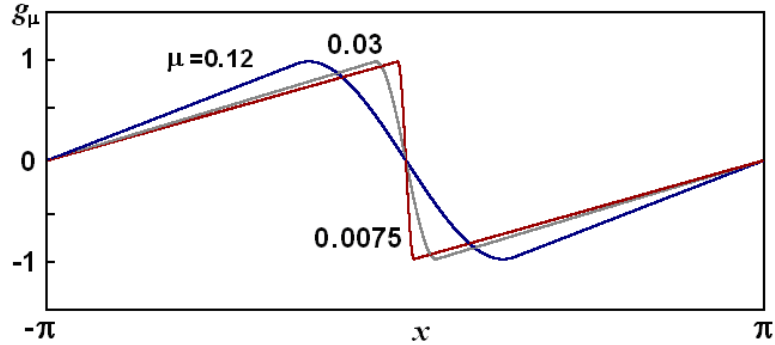

Fig.6. Functions of the family $g_\mu(x)$ for different values of the smoothing parameter.

Figure 7 shows attractor in the extended phase space for the continuous-time system (2) with a smoothed sawtooth function, where the smoothing parameter is set equal to 0.03. It can be compared with the portrait of the Belykh attractor in Fig. 5. (To see the differences, pay attention to the central part of the picture.) Due to the periodicity of the extended phase space over the time coordinate, the picture should be imagined as repetitive with period 1 along the vertical axis.

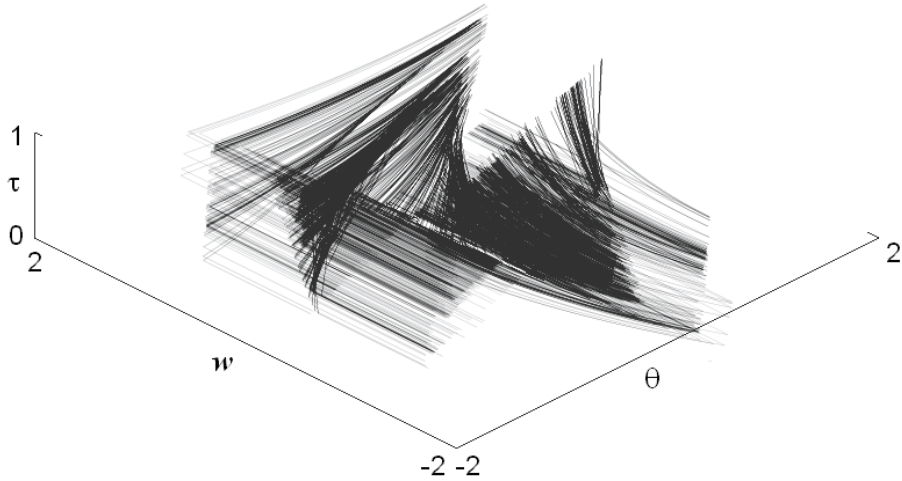

Fig.7. Attractor of the system (2) in the extended phase space with the smoothed function $g_\mu(\theta)$ for µ=0.03. The remaining parameters are the same as in Fig.5.

Figure 8 shows two-dimensional portraits of the attractors of the discrete-time system (4) with the function (14) for different values of the smoothing parameter µ.

As can be seen in the diagrams *a, b, c*, when smoothing the sawtooth function, the arrangement of the Belykh attractor is modified: the structure is complemented by appearance of "filaments" connecting previously separated oblique "upper" and "lower" parts of the attractor.

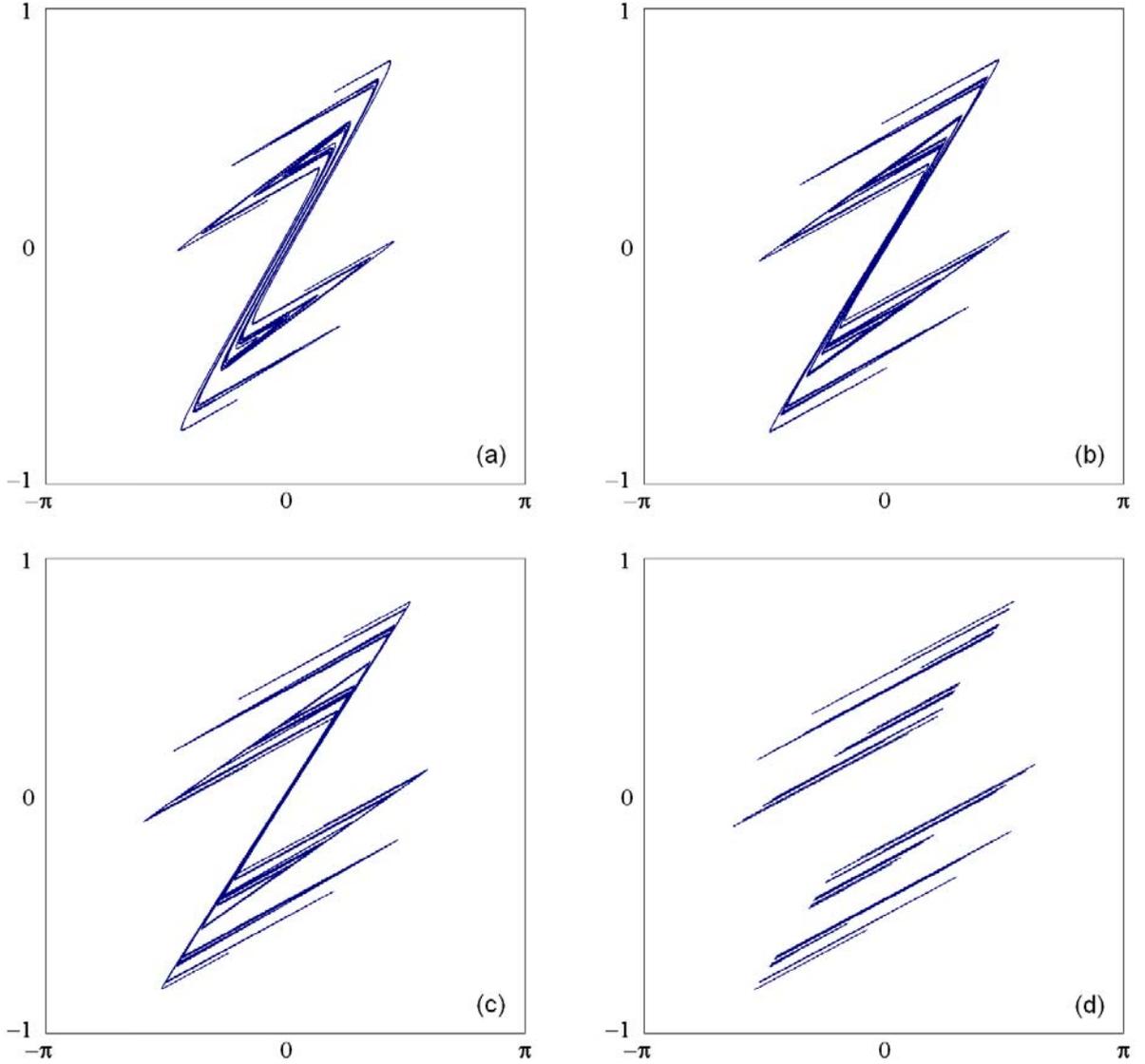

Fig.8. Attractor of the map with smoothed function given by (14) with μ= 0.03 (a), 0.015 (b) and 0.0075 (c), and the Belykh attractor (d). Parameters are $\lambda = 0.48$, $a = 1.2$, $\Delta = 0$.

Figure 9 shows a graph of Lyapunov exponents calculated numerically using the standard algorithm [27, 28, 14], depending on the smoothing parameter μ. The appearance of drops on the graph of the largest exponent into the negative region are associated with "windows" of periodic dynamics arising due to a possibility of visiting the region of smoothing of the sawtooth function by a trajectory on the attractor. This indicates violation of the quasi-hyperbolic nature of the attractor. In calculations with better resolution in the parameter, and with extended time interval for estimating Lyapunov exponents, the number of distinguishable "windows" increases.

Completing the presentation, it is appropriate to recall that one of the characteristic properties of the Zaslavsky map with a smooth function $g(x)$ consists in presence of regions of periodic regimes of dynamics in the parameter plane called the Arnold tongues [14]. It seems interesting to discuss this in the context of the picture described above.

Figure 10 shows charts of dynamical regimes of the Zaslavsky map (4) with the function (14) for different values of the smoothing parameter in the plane of the detuning parameters Δ and the intensity of the kicks *a*. The procedure consisted in scanning the grid nodes with a certain step in two parameters. At each point, about $10^3$ iterations of the map are performed together with the linearized mapping to estimate the Lyapunov exponent. Basing on the last iteration steps, an analysis is performed for the presence or absence of a period of repetition of the states (in the sense of coincidence of the angular coordinate modulo 2π) with a certain initially assumed level of the allowable error. If a periodicity is detected, the corresponding pixel in the diagram is indicated by a certain color, determined by the period of repetition of states, and the routine passes to the next point on the parameter plane. Here it is reasonable to specify the initial conditions at the new point as a final state of the iterations at the previous point ("scanning with inheritance"), which accelerates convergence to the sustained dynamical mode. Black color on the chart corresponds to non-periodic dynamics. White color indicates areas of chaos, diagnosed by a positive value of the largest Lyapunov exponent.

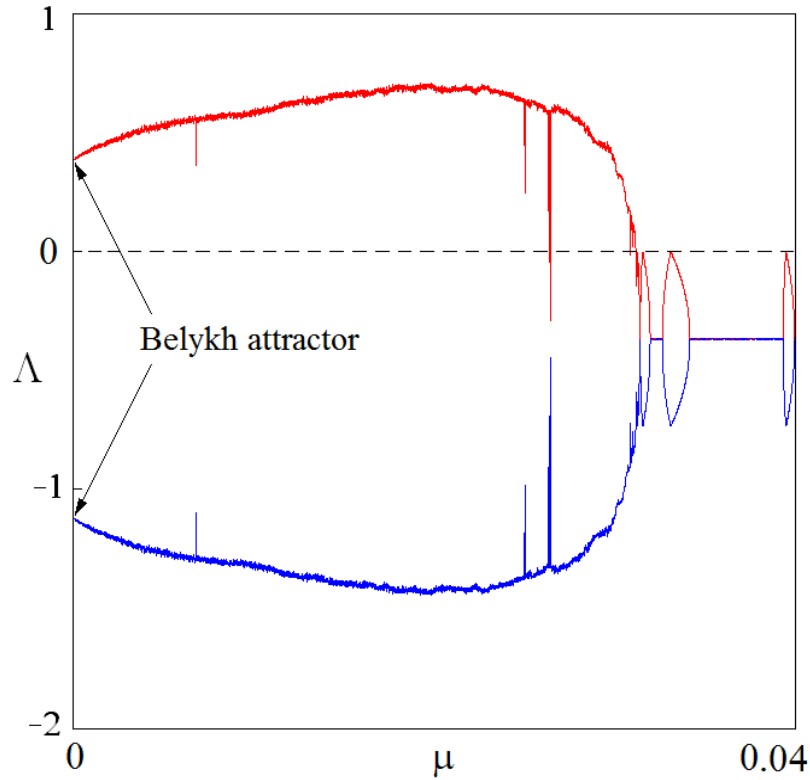

Fig.9. Lyapunov exponents depending on the smoothing parameter μ for the map with function (14); the remaining parameters are λ = 0.48, a = 1, Δ = 0.

Periodic behavior is observed in the regions in the form of characteristic tongues (Arnold's tongues). Periodic regimes inside Arnold's tongues are interpreted as synchronization of the system by external periodic driving. The presence of a large number of the tongues suggests that the synchronization can be provided at various harmonics and subharmonics of the external driving. In fact, the picture shows only the widest tongues. While the parameter $a$ is small enough, there is place yet for quasi-periodic regimes between the tongues. In the region of large values of $a$, within the Arnold tongues one can observe a complex picture, including cusp points, fold lines and period doubling lines. When moving inside an Arnold tongue in the general direction of increase of $a$, one can observe emergence of chaos through the period doubling cascade. The calculations show that the Arnold tongues, which take place for present for the smoothed function $g_\mu(x)$ decrease in size and vanish in the limit case of the discontinuous sawtooth function, μ→0.

Figure 11 shows the plots of two Lyapunov exponents for the Zaslavsky map versus the detuning parameter, with a constant parameter $a$, for three different values of the smoothing parameter μ. These graphs correspond to a route on the parameter plane in Fig. 10 along the horizontal line $a$=1. There again we can observe "dips", the windows of periodicity that correspond on the chart of regimes to narrow bands of regular dynamics stretched out into the chaos area. In regions where there are no distinguishable "dips" the Lyapunov exponents are close to the values corresponding to the Belykh attractor (12). With decrease in the smoothing parameter, the number of distinguishable "windows" with the resolution adopted in the calculations becomes smaller, and in the limit μ → 0 they obviously disappear.

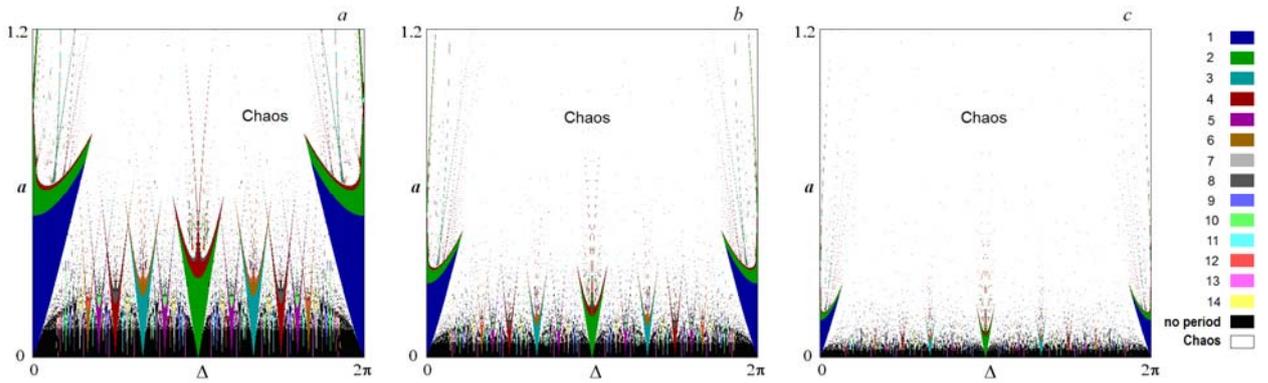

Fig.10. A picture of Arnold's tongues on the parameter plane of the map with the function (14) depending on the smoothing parameter: μ=0.03 (a), 0.015 (b) and 0.0075 (c). The parameter λ equals 0.48. White color corresponds to chaos (positive Lyapunov exponent), and black to quasi-periodic dynamics. Legend for the colors associated with the periods is shown to the right.

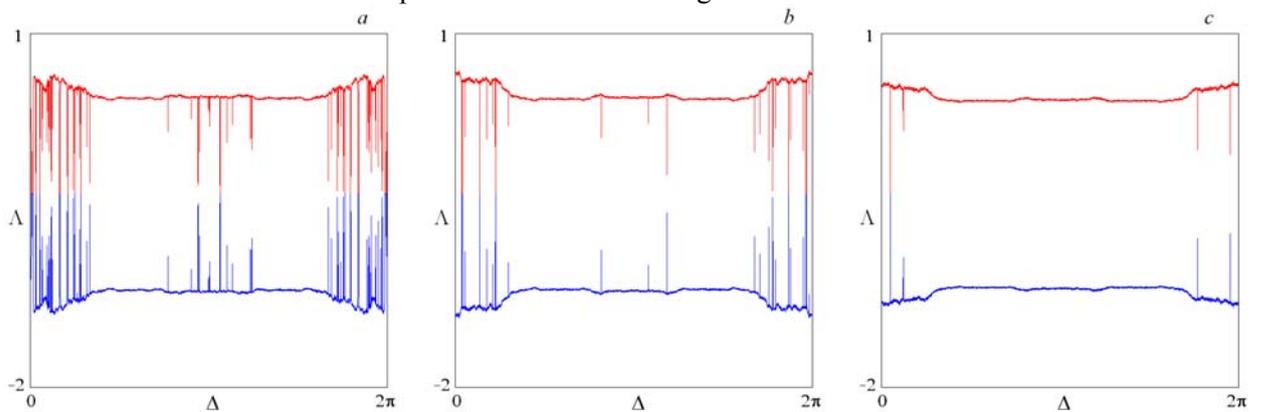

Fig.11. Lyapunov exponents, depending on the detuning parameter Δ for the map with the function (14) for different values of the smoothing parameter: μ=0.03 (a), 0.015 (b) and 0.0075 (c). The remaining parameters are λ = 0.48, a = 1.

## Conclusion

Attractor of Belykh is a quasi-hyperbolic attractor of a system with evolution operator defined using a discontinuous function.

Attractors of this type occur in radiophysical context. Here, for concreteness, it is considered in Zaslavsky map describing a dissipative rotator with periodic kicks, with a sawtooth form of the dependence of the intensity of kicks on the angular coordinate. The derivations are presented bringing the equations to the standard form of the Belykh map. Illustrations of the dynamical behavior of the system are provided. It is shown that the smoothing of the sawtooth function leads to destruction of the quasi-hyperbolic nature of the attractor and to appearance of phenomena intrinsic to the quasiattractor, namely, windows of regularity in parameter dependence of the dynamics, and dips in the graphs of Lyapunov exponent. However, with a small-scale smoothing, there are regions in the parameter space, where these windows are practically indistinguishable and in experimental implementations of the system will be effectively masked by noise. In these cases, electronic devices with this type of attractors apparently can be used as generators of practically robust chaos, ignoring the difference from the quasi-hyperbolic situation.

*This work was supported by the grant of the Russian Science Foundation No. 17-12-01008.*